\documentclass[showpacs,epsfig,preprintnumbers,amsmath,amssymb,nofootinbib]{revtex4-1}
\usepackage{graphicx}
\usepackage{dcolumn}
\usepackage{bm}
\usepackage{subfigure}
\usepackage{comment}
\usepackage{epstopdf}
\begin{document}
\def\d{{\rm d}}
\def\Epos{E_{\rm pos}}
\def\ap{\approx}
\def\eff{{\rm eft}}
\def\L{{\cal L}}
\newcommand{\vev}[1]{\langle {#1}\rangle}
\newcommand{\CL}   {C.L.}
\newcommand{\dof}  {d.o.f.}
\newcommand{\eVq}  {\text{EA}^2}
\newcommand{\Sol}  {\textsc{sol}}
\newcommand{\SlKm} {\textsc{sol+kam}}
\newcommand{\Atm}  {\textsc{atm}}
\newcommand{\Chooz}{\textsc{chooz}}
\newcommand{\Dms}  {\Delta m^2_\Sol}
\newcommand{\Dma}  {\Delta m^2_\Atm}
\newcommand{\Dcq}  {\Delta\chi^2}
\newcommand{\nbb}{$\beta\beta_{0\nu}$ }
\newcommand {\be}{\begin{equation}}
\newcommand {\ee}{\end{equation}}
\newcommand {\ba}{\begin{eqnarray}}
\newcommand {\ea}{\end{eqnarray}}
\def\VEV#1{\left\langle #1\right\rangle}
\let\vev\VEV
\def\e6{E(6)}
\def\10{SO(10)}
\def\21{SA(2) $\otimes$ U(1) }
\def\321{$\mathrm{SU(3) \otimes SU(2) \otimes U(1)}$ }
\def\lr{SA(2)$_L \otimes$ SA(2)$_R \otimes$ U(1)}
\def\422{SA(4) $\otimes$ SA(2) $\otimes$ SA(2)}
\newcommand{\AHEP}{%
School of physics, Institute for Research in Fundamental Sciences
(IPM)\\P.O.Box 19395-5531, Tehran, Iran\\

  }
\newcommand{\Tehran}{%
School of physics, Institute for Research in Fundamental Sciences
(IPM)
\\
P.O.Box 19395-5531, Tehran, Iran}
\def\roughly#1{\mathrel{\raise.3ex\hbox{$#1$\kern-.75em
      \lower1ex\hbox{$\sim$}}}} \def\lsim{\roughly<}
\def\gsim{\roughly>}
\def\ltap{\raisebox{-.4ex}{\rlap{$\sim$}} \raisebox{.4ex}{$<$}}
\def\gtap{\raisebox{-.4ex}{\rlap{$\sim$}} \raisebox{.4ex}{$>$}}
\def\lsim{\raise0.3ex\hbox{$\;<$\kern-0.75em\raise-1.1ex\hbox{$\sim\;$}}}
\def\gsim{\raise0.3ex\hbox{$\;>$\kern-0.75em\raise-1.1ex\hbox{$\sim\;$}}}


\title{Shedding light  on LMA-Dark solar neutrino solution by medium baseline reactor experiments: JUNO and RENO-50 }

\date{\today}
\author{P. Bakhti}\email{pouya\_bakhti@ipm.ir}
\author{Y. Farzan}\email{yasaman@theory.ipm.ac.ir}
\affiliation{\Tehran}
\begin{abstract}
In the presence of Non-Standard neutral current Interactions (NSI) a
new solution to solar neutrino anomaly with $\cos 2\theta_{12}<0$
appears. We investigate how this solution can be tested by upcoming
intermediate baseline reactor experiments, JUNO and RENO-50. We point out a degeneracy between the two solutions when both hierarchy and the $\theta_{12}$ octant are flipped. We then comment on how this degeneracy
can be partially lifted by long baseline experiments sensitive to matter effects such as the NOvA experiment. 
\end{abstract}
{\keywords{Neutrino, Leptonic CP Violation, Leptonic Unitary
Triangle, Beta Beam}}
\maketitle
\section{Introduction}
Within the Standard Model (SM) the neutral current interactions are
flavor diagonal and universal for all three flavors.  However, most
beyond SM mechanisms dealing with flavor predict a correction to
neutrino  interaction terms which violate flavor universality and
conservation. Examples of such models include $R$-parity violating
supersymmetry, grand unification, AMEND model \cite{AMEND}, extra
$U(1)'$ gauge models, left-right symmetric models and various seesaw
models (for a review see \cite{Tommy}). The non-standard  neutral
current interaction of neutrinos  can be in general formulated by an
effective dimension six operator as \be \label{NSI}
\mathcal{L}_{NSI}=-2\sqrt{2}G_F\epsilon_{\alpha \beta}^{f P}
(\bar{\nu}_\alpha \gamma^\mu L\nu_\beta)(\bar{f}\gamma_\mu P ~f) \ee
where $f$ is the matter field ($u, \ d$ or $e$), $P$ is the
chirality projection matrix and $\epsilon_{\alpha \beta}^{f P}$ is a
dimensionless matrix describing the deviation from the standard
model. For neutrino oscillation, only the ``vector'' part of the
interaction operator is relevant so it is convenient to define
$$ \epsilon_{\alpha \beta}^{f }\equiv \epsilon_{\alpha \beta}^{f L}+\epsilon_{\alpha \beta}^{f R}.$$
Effects of Lagrangian (\ref{NSI}) on neutrino oscillation have been
extensively studied  in the literature. In particular in
\cite{Tortola}, it is shown that in the presence of a deviation from
universality ({\it i.e.,} $|\epsilon_{ee}^f -\epsilon_{\mu
\mu}^f|,|\epsilon_{ee}^f -\epsilon_{\tau \tau}^f|\ne 0$ with
$f=u,d$), another solution with $\cos(2\theta_{12})<0$ for solar and
KamLAND data exists. This solution is known as LMA-Dark solution.
Recent studies show that this new solution survives combining all
the available data on oscillation \cite{Maltoni}. In fact in
presence of Non-Standard Interactions (NSI), the fit to solar data
is slightly better as in the presence of NSI, the upturn of the
spectrum at low energy predicted by the standard LMA solution
without NSI can be suppressed, leading to a better agreement with
the data \cite{Tortola}. The NSI can also affect other observable
quantities such as the invisible decay width of the $Z$ boson (at
one-loop) or neutrino  scattering off matter. All relevant bounds
have extensively been studied \cite{Tommy,Biggio,Davidson}. The
bound from the CHARM scattering experiment combined with the NuTeV
results rule out a part of the parameter space relevant for the
LMA-Dark solution ({\it i.e.,} $0.9<|\epsilon_{ee}^d-\epsilon_{\mu
\mu}^d|<0.8$ at 90 \% C.L.) \cite{Davidson,Maltoni}. However, the
LMA-Dark solution is not completely ruled out and needs further
investigation.

Recently, two intermediate baseline neutrino experiments JUNO and
RENO-50 have been proposed to be built in China and south Korea,
respectively. Determining the neutrino mass hierarchy ({\it i.e.,}
normal vs inverted) and precision measurement of the solar mixing
parameters $\theta_{12}$ and $\Delta m_{21}^2$  are the prime goals
of these experiments
\cite{Ge:2012wj,first,Li,Learned,Blen,Evslin,500,ematters,Ciuffoli}
(see also \cite{piai}). Recently, we have shown that the data from
these two experiments can also be employed to probe the superlight
sterile neutrino scenario \cite{us}. Refs. \cite{Tahir,Tommy2} study
the effects of charged current NSI at detector and source. The aim
of the present paper is to show that the medium baseline experiments
can help to probe the LMA-Dark solution  for which
$\theta_{12}>\pi/4$. We find a degeneracy between solutions when signs of both $\cos 2 \theta_{12}$ and $\Delta m_{31}^2$ are simultaneously flipped and then discuss the possibility of lifting this degeneracy. 

The rest of the paper is organized as follows.  In sect.~\ref{Osc},
we review the oscillation probability and discuss how the medium
baseline reactor experiments distinguish between the LMA and
LMA-Dark solutions. In sect.~\ref{JUNO-RENO-50}, we describe the
JUNO and RENO-50 experiments and list the background. In
sect.~\ref{Results}, we present our numerical results obtained via
the GloBES software \cite{Globes}. 
In sect.~\ref{degeneracy}, we discuss the degeneracy in more detail and examine the possibility of lifting it.
In sect.~\ref{Con}, we summarize our results.
\section{Oscillation probability \label{Osc}}
The energy of the reactor neutrinos are of order of MeV so in the
leading order, the matter effects can be neglected in the
propagation of these neutrinos in the earth ({\it i.e.,} $\Delta
m_{21}^2/E_\nu \gg \sqrt{2} G_F N_e$). As a result, the effect of
neutral current NSI in Eq.~(\ref{NSI}) on neutrino propagation can
also be neglected. In fact,  Refs. \cite{Tahir,Tommy2} focus on the
charged current NSI that affect production and detection [{\it
i.e.,} $(\bar{d}\gamma^\mu P~u)( \bar{e} \gamma_\mu L
\nu_{\mu(\tau)})$]. Neutral current interaction of type (\ref{NSI})
cannot affect the production and detection either. At first sight,
it seems counterintuitive that reactor neutrinos help us to probe
the impact of neutral current NSI. Notice however that we are
proposing to determine $\cos 2\theta_{12}$ rather than constraining
the NSI parameters, $\epsilon_{\alpha \beta}^f$. Neglecting the
matter effects, one can write \be \label{Pee} P(\bar{\nu}_e \to
\bar{\nu}_e) =\left| |U_{e1}|^2 +|U_{e2}|^2e^{i
\Delta_{21}}+|U_{e3}|^2e^{i \Delta_{31}}\right|^2= \left|
c_{12}^2c_{13}^2+ s_{12}^2 c_{13}^2 e^{i\Delta_{21}}+s_{13}^2 e^{i
\Delta_{31}}\right|^2= \ee
$$c_{13}^4 (1-\sin^2 2\theta_{12}
\sin^2\frac{\Delta_{21}}{2})+s_{13}^4+2s_{13}^2c_{13}^2[\cos
\Delta_{31}(c_{12}^2+s_{12}^2 \cos \Delta_{21})+s_{12}^2\sin
\Delta_{31}\sin \Delta_{21}]$$
where $\Delta_{ij}=\Delta m_{ij}^2L/(2E_\nu)$ in which $L$ is the
baseline. For short baseline reactor experiments such as Daya Bay,
RENO or (double-)CHOOZ, we can set $\Delta_{21}\simeq 0$ so the
sensitivity to $\theta_{12}$ is lost altogether. At KamLAND,
$\Delta_{21}$ is sizeable but the oscillatory modes given by
$\Delta_{31}$ are averaged out so KamLAND is only sensitive to
$\sin^2 2\theta_{12}$ which cannot distinguish  between the two
solutions with $\theta_{12}>\pi/4$ and $\theta_{12}<\pi/4$. To
distinguish between the standard LMA and LMA-dark solutions the
experiment should be sensitive to the last terms in Eq. (\ref{Pee})
given by $\cos\Delta_{31} \cos\Delta_{21}$ and  $\sin\Delta_{31}
\sin\Delta_{21}$. The JUNO and RENO-50 experiments are proposed to
resolve these terms as the term given by $\sin\Delta_{31}
\sin\Delta_{21}$ is the one sensitive to sign$( \Delta_{31})$ and
hence the mass hierarchy scheme. In principle, by studying the
energy spectrum of the events,   we can resolve these terms and
extract  their amplitude and sign. Thus, we can discriminate  
 between the standard LMA and non-standard LMA-Dark
solutions. However, it  is a non-trivial question to determine
whether this can in principle be possible taking into account the
realistic uncertainties. In the rest of the paper, we try to address
this question. Before proceeding further notice that $P(\bar{\nu}_e
\to \bar{\nu}_e )$  in Eq. (\ref{Pee}) is invariant under \be
\label{transform} s_{12} \leftrightarrow c_{12}~({\it i.e.,} ~
\theta_{12}\to \frac{\pi}{2}-\theta_{12}) \ \ \ {\rm and} \ \ \
\Delta_{31} \to -\Delta_{31}+\Delta_{21} \ . \ee  In other words, as
far as we neglect matter effects, there is a degeneracy when we
simultaneously flip hierarchy (NH$\leftrightarrow$IH) and flip
between the LMA and LMA-Dark solutions. We will discuss more about
this degeneracy in sect. IV and in sect \ref{degeneracy}, we will generalize this symmetry to include matter effects.
\section{ JUNO and RENO-50 experiments \label{JUNO-RENO-50}}

The Juno and RENO-50 experiments with baselines of $L\sim 50$~km are
scheduled to become ready for data taking in 2020 \cite{reno50}. The
detectors will use liquid scintillator technique with an energy
resolution of $$\frac{\delta E_\nu}{E_\nu} \simeq 3\%
\times(\frac{E_\nu}{ {\rm MeV}})^{1/2}.$$   Ref. \cite{Learned}
enumerates  the following backgrounds as the dominant ones  (i)
accidental background; (ii) $^{13} C(\alpha,n) ^{16}O$ background
and (iii) Geoneutrino background. For the spectrum of these sources
of background  and their normalization we use values and description
respectively in \cite{bs} and  in \cite{Learned}.  However, as shown
in recent paper \cite{Lithium}, the background caused by $^9Li$ from
cosmic muon interaction will be dominant.  We take  10000 and 5000
fake neutrino signals due to $^9Li$ at respectively JUNO and RENO-50
and assume a spectrum of shape given in \cite{talk} for them. The
reason why the cosmic muon induced $^9 Li$ background  is
substantially less for RENO-50 than that for JUNO is the deeper
location of RENO-50 detector and therefore better shielding from
cosmic muons. Notice that the normalization we take for $^9Li$
background is relatively conservative. Reconstructing the muon
tracks and using a smart veto, the background can be reduced down to
half the assumed value \cite{private}.

We divide the energy range between 1.8 MeV to 8 MeV to 350 bin of
size 17.7 keV in our analysis. We take the energy calibration error
equal to 3 \%. Let us now describe the features specific for each
experiment one by one.
\paragraph{The JUNO experiment:} JUNO will be located at a distance of  52 km far
from Yangjiang and Taishan reactor complexes  with a combined power
of 36 GW \cite{Li}. JUNO will also receive neutrino flux from the
existing Daya Bay and planned Huizhou reactors respectively located
215 km and 265 km far from it.    We take the flux normalization
uncertainty to be 5 \%. The scintillator detector will have a
fiducial mass of 20 kton. A
 list of reactor distances and powers  can be found in \cite{Li}. To simplify computation, in our
 numerical analysis we
 combine the reactor cores whose distance to detector are close to
 each other. Table 1 summarizes the powers and baselines that we
 take in our analysis.

\begin{table}[t]
\begin{center}
\begin{tabular}{ |c|c|c|c|c|}
    \hline
    reactor core & 1 & 2 & 3 & 4\\ \hline \hline
    Baseline (km) & 52.17 & 52.36 & 52.58 & 52.80  \\ \hline
    Power (GW) & 10.4 & 7.5 & 7.5 & 10.4 \\ \hline

    \end{tabular}
\caption{ Baselines and powers of reactor cores  taken for the JUNO
experiment.} \label{tab:tab1}
\end{center}
\end{table}

\paragraph{The RENO-50 experiment} The RENO-50 setup is an upgrade of
the current RENO experiment using the neutrino flux from the same
reactors with a total power of 16.4 GW. The current detector will be
used as near detector reducing the flux uncertainty down to 0.3 \%
\cite{Seo}. The far detector with a fiducial mass of 18 kton will be
located 47 km away.


The potential of reactor neutrino experiments with a baseline of
$\sim $ 50 km  for determining  the neutrino mass ordering has been
extensively studied in the literature
\cite{piai,Blen,Evslin,Li,Ciuffoli}. The main goal of JUNO and
RENO-50 experiments is determining the sign of $\Delta m_{31}^2$.
 It is
shown that in order to determine sgn($\Delta m_{31}^2$), the
difference between the distances of different reactor cores
contributing to the flux of the detector should be less than
$\mathcal{O}(500)$ meters \cite{500,Evslin,Li}. Considering this
restriction, the best location for JUNO is found to be at a 52 km
distance from Yangjiang and Taishan reactor complexes
\cite{first,Li}.  Like the case of determining the hierarchy, we
expect the distribution of reactor sources to reduce the sensitivity
to ${\rm sign}(\cos 2\theta_{12})$ because the distribution of the
sources lead to average out of the effects of the  oscillatory terms
given by $\Delta_{31}$. Although the matter effects are subdominant,
in the numerical analysis we take them into account.

From Eq. (\ref{Pee}), we observe that the terms sensitive to
sign($\cos 2 \theta_{12}$) are suppressed by $s_{13}^2 \sim 2.5 \%$.
Thus, at first glance it seems that an uncertainty of 3 \% or larger
in the shape of the initial energy spectrum can wash out the
sensitivity to sign($\cos 2 \theta_{12}$) as well as the sensitivity
to sign($\Delta m^2_{13}$).  In fact, the uncertainty in the shape
of the initial energy spectrum at source is at the level of $O(3
\%)$ \cite{MuellerANDHuber}. However as we discuss below, the
effects of this uncertainty can be safely neglected. Let us denote
the uncertainty in the shape of the initial energy spectrum at
energy bin ``$i$'' by $\Delta \alpha_i$. We take into account the
effect of this uncertainty by pull method, defining \be
\label{deviation} \chi^2={\rm Min}|_{\theta_{pull}, \alpha_i}\left[
\sum_i
\frac{[N_i(\theta_0,\bar{\theta}_{pull})-N_i(\theta,{\theta}_{pull})(1+\alpha_i)]^2}{N_i(\theta_0,\bar{\theta}_{pull})}+
\sum_i \frac{\alpha_i^2}{(\Delta
\alpha_i)^2}+\frac{(\theta_{pull}-\bar{\theta}_{pull})^2}{(\Delta
\theta_{pull})^2}\right],\ee where  $\alpha_i$ is the pull parameter
taking care of the uncertainty in the initial spectrum at bin $i$.
$\theta_{pull}$ collectively denotes pull parameters other than
$\alpha_i$ which have true values collectively denoted by
$\bar{\theta}_{pull}$ and  uncertainties collectively denoted by
$\Delta \theta_{pull}$. $\theta$ and $\theta_0$ are respectively the
fit parameter and its true value. $N_i$ is the number of events at
bin $i$. To calculate the deviation, we minimize over each
$\alpha_i$ as well as over all $\theta_{pull}$. It is
straightforward to show that as long as \be \label{condition} N_i
(\Delta \alpha_i)^2 \ll 1,\ee we can neglect the effects of $\Delta
\alpha_i$ in evaluating $\chi^2$. Considering Fig (13) of Ref.
\cite{white} and uncertainties found in \cite{MuellerANDHuber}, we
observe that even with spectrum divided into bins of size 17.7 keV,
the condition in (\ref{condition}) is fulfilled  so the present
uncertainty in the shape of the spectrum will not be a major
limitation for extracting sign($\cos (2 \theta_{12})$) and/or
sign($\Delta m_{31}^2)$.

To carry out our analysis, we employ the GLoBES software
\cite{Globes}.  We use  the reactor neutrino energy spectrum  and
neutrino cross section that are respectively  given in
\cite{Murayama:2000iq,Eguchi:2002dm} and \cite{Vogel:1999zy}. For
neutrino mass and mixing parameters, we take the best fit values
listed in \cite{GonzalezGarcia:2012sz}.  We assume an uncertainty of
6\%  both in  $\theta_{13}$ and in $\Delta m_{21}^2$.
We use the pull-method to treat the
uncertainties.
\section{ Numerical results \label{Results}}

\begin{figure}
\begin{center}
\subfigure[]{\includegraphics[width=0.49\textwidth]{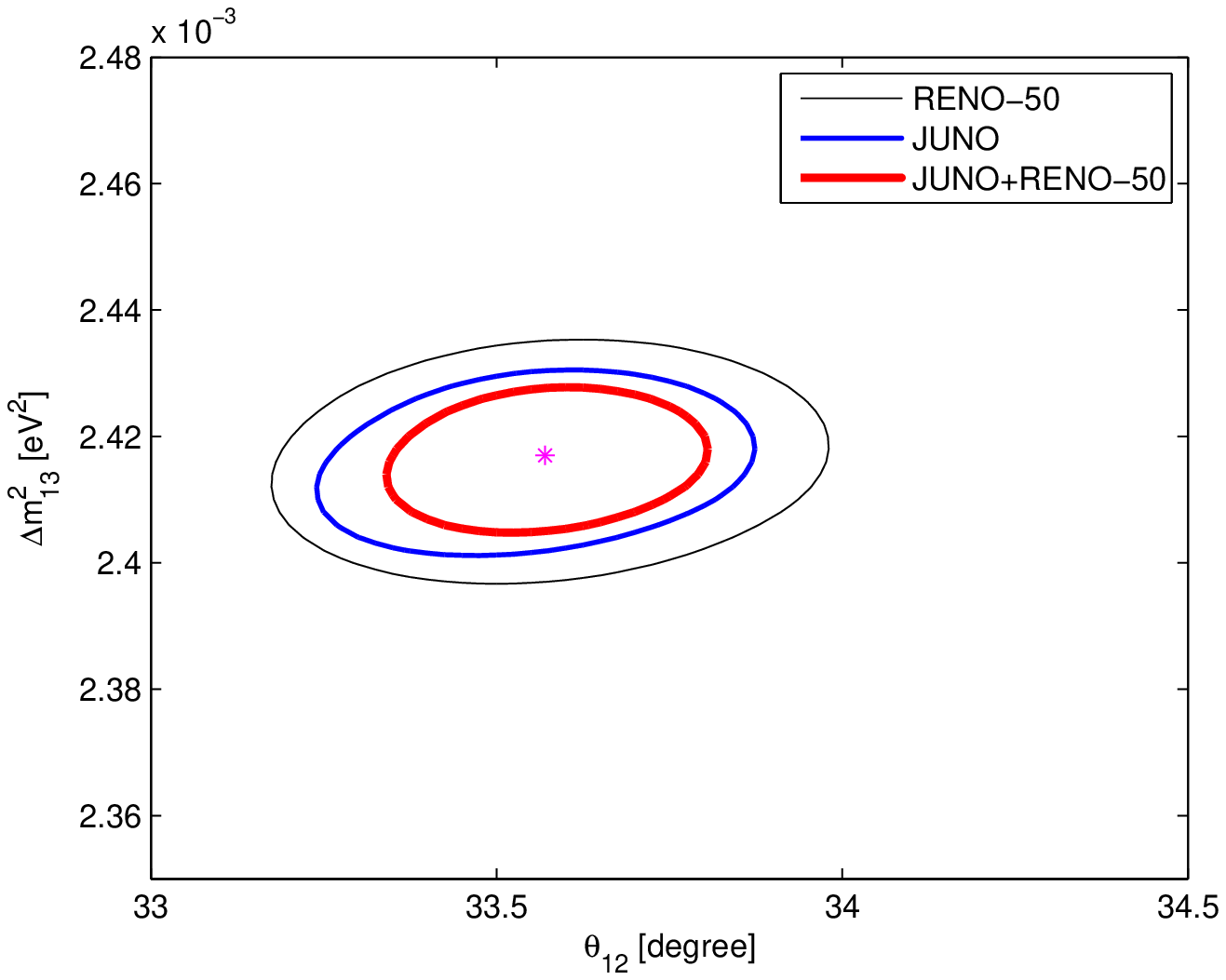}}
\subfigure[]{\includegraphics[width=0.49\textwidth]{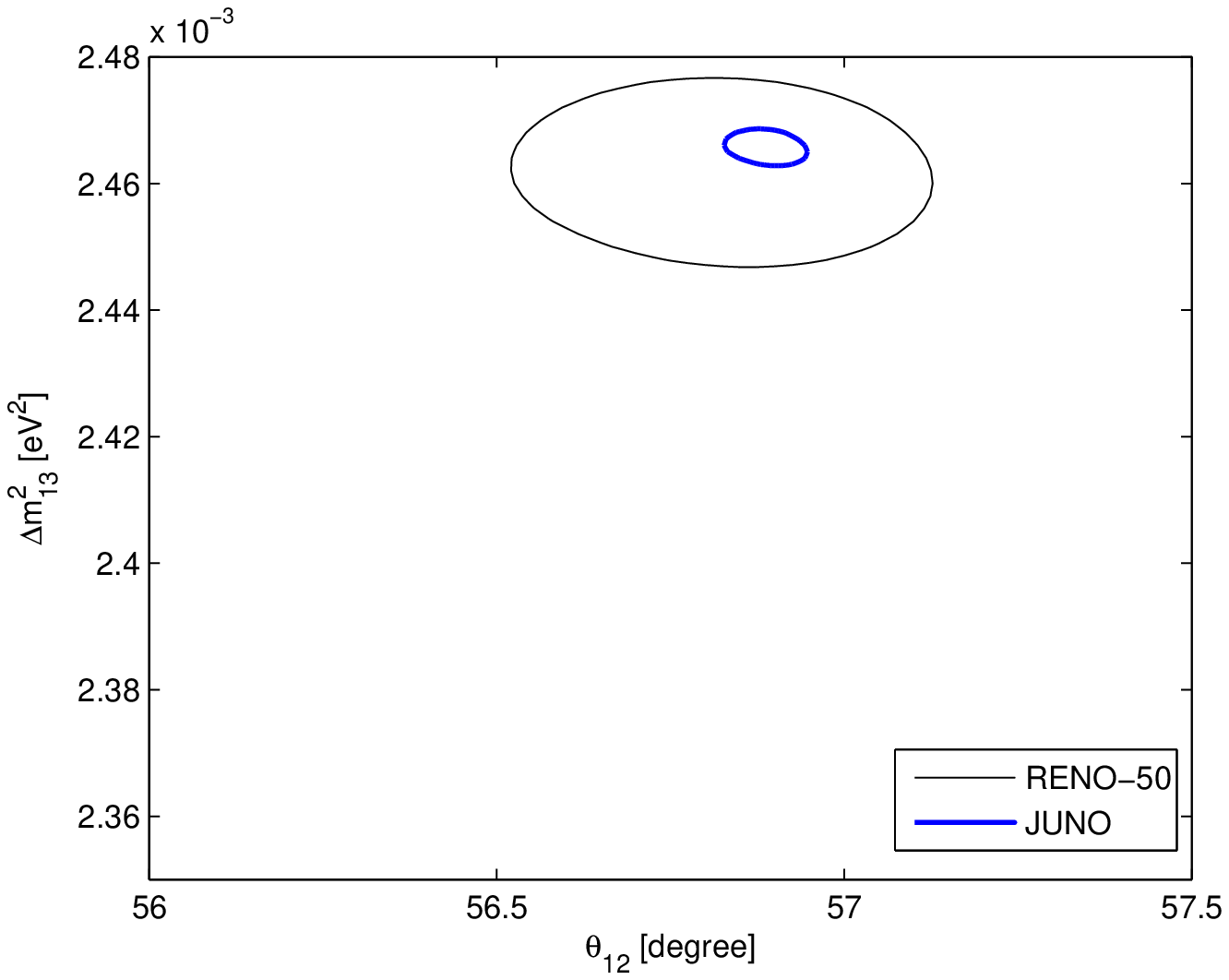}}
\subfigure[]{\includegraphics[width=0.49\textwidth]{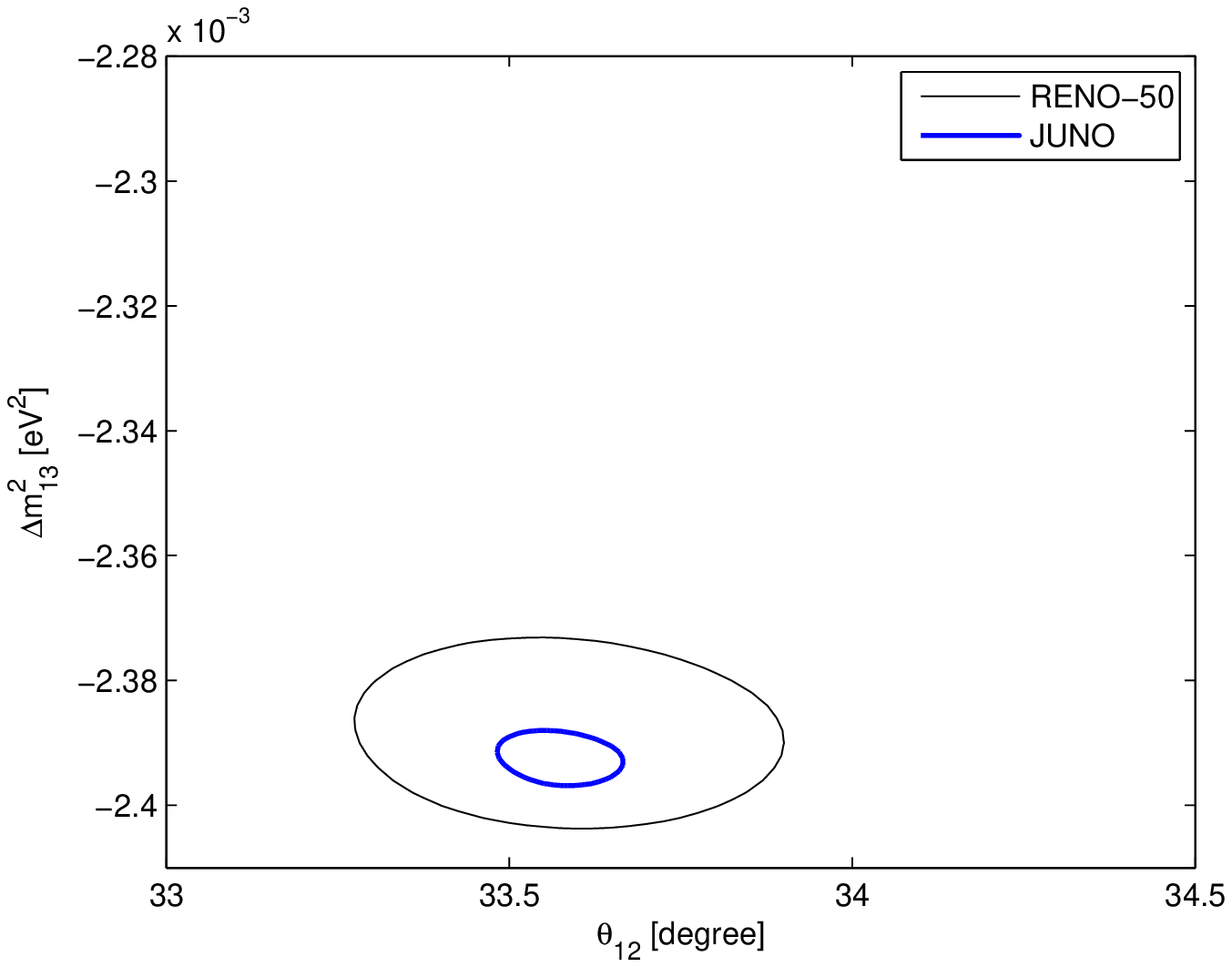}}
\subfigure[]{\includegraphics[width=0.49\textwidth]{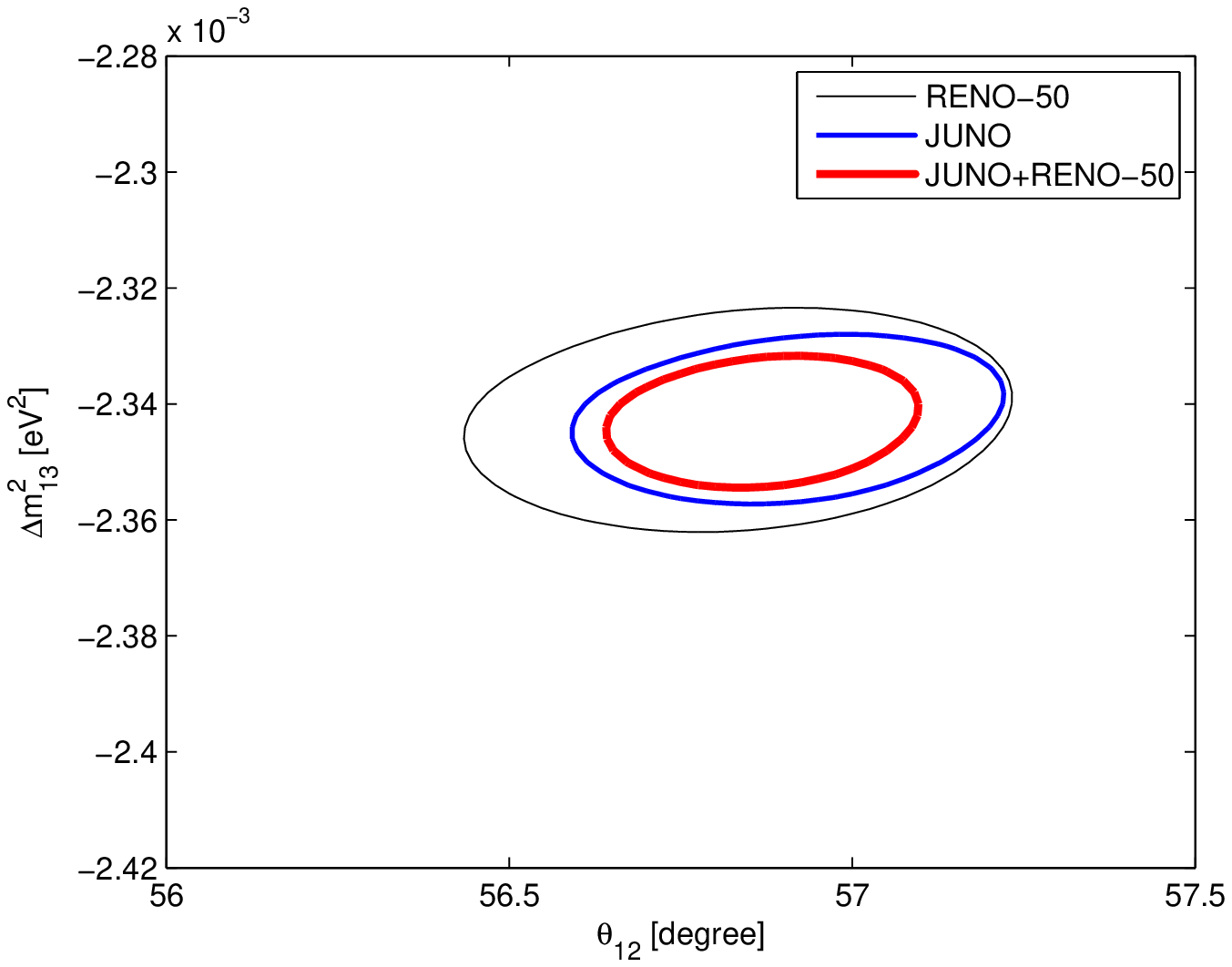}}
\end{center}
\vspace{2cm} \caption[]{Allowed region at 3 $\sigma$ C.L. after 5
years of data taking by RENO-50 and JUNO. The true values of the
neutrino parameters, marked with a star in Fig. (a), are taken to be
$\Delta m_{31}^2=2.417\times10^{-3}~{\rm eV}^2$,
$\theta_{12}=33.57^\circ$, $\Delta m_{21}^2=(7.45\pm
0.45)\times10^{-5}~{\rm eV}^2$ and $\theta_{13}=(8.75\pm
0.5)^\circ$. The upper (lower) panels show the allowed region for
normal (inverted) hierarchy and left (right) panels show LMA
(LMA-Dark) solution for $\theta_{12}$.}

\label{Contour-bright}

\end{figure}

\begin{figure}
\begin{center}
\subfigure[]{\includegraphics[width=0.49\textwidth]{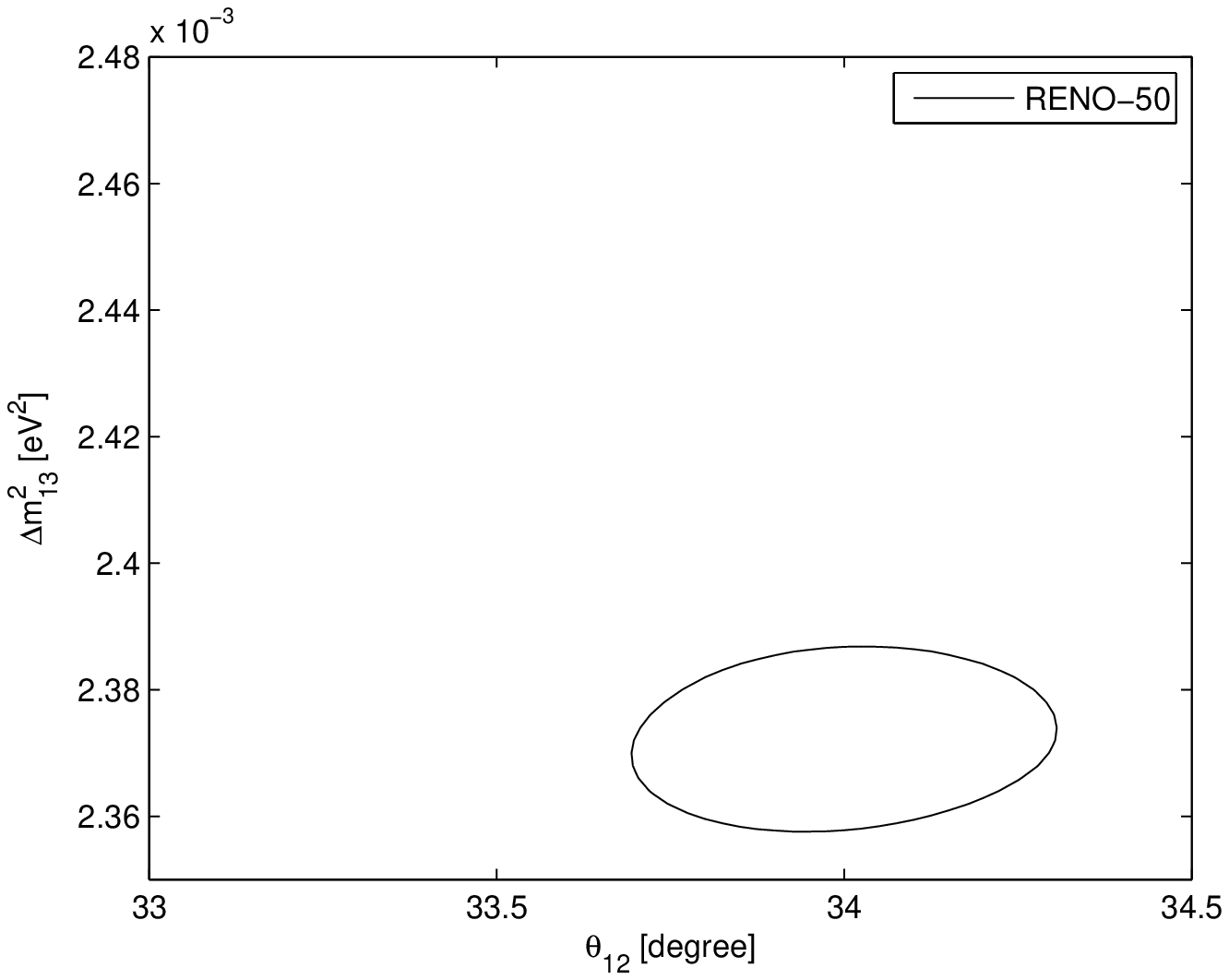}}
\subfigure[]{\includegraphics[width=0.49\textwidth]{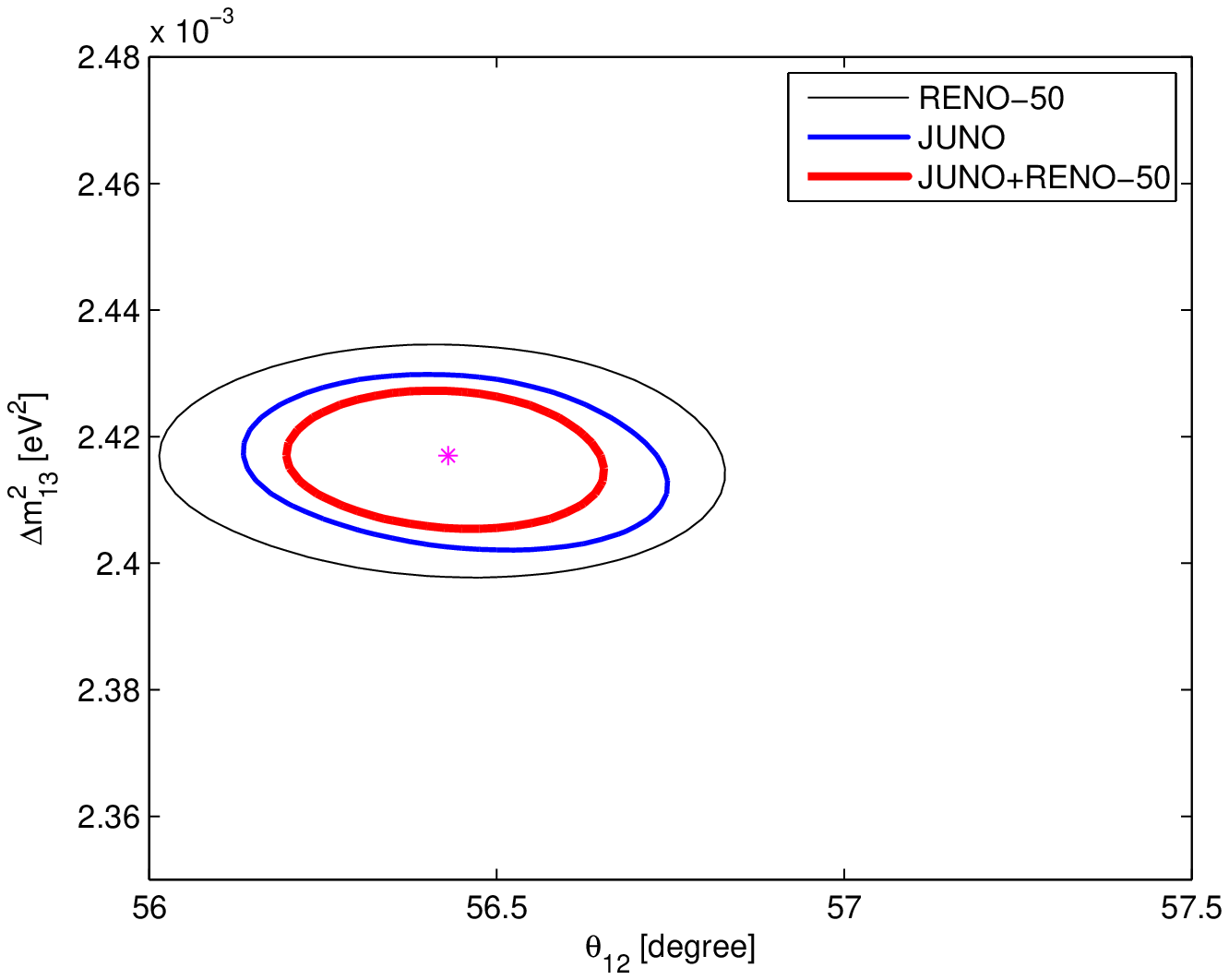}}
\subfigure[]{\includegraphics[width=0.49\textwidth]{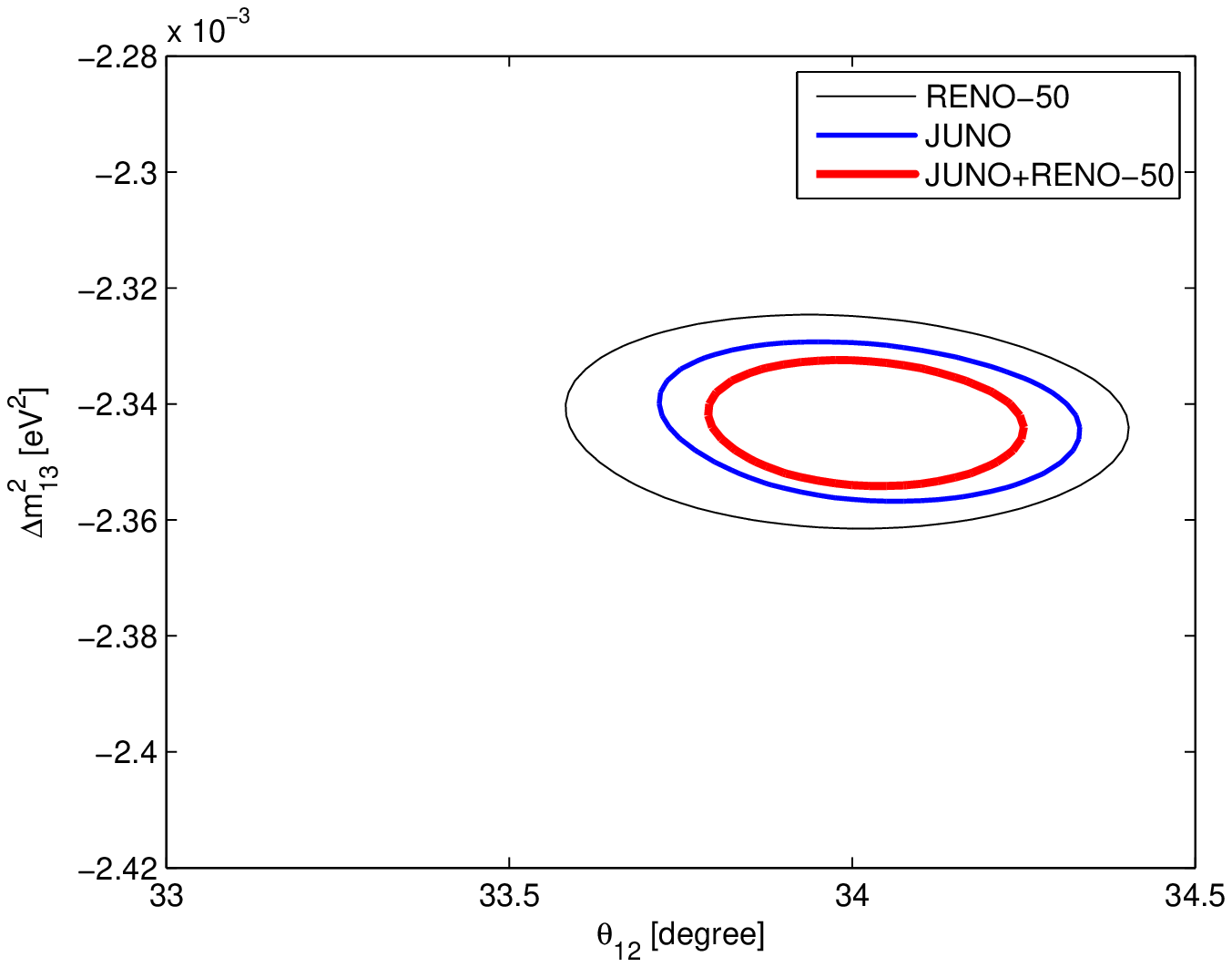}}
\subfigure[]{\includegraphics[width=0.49\textwidth]{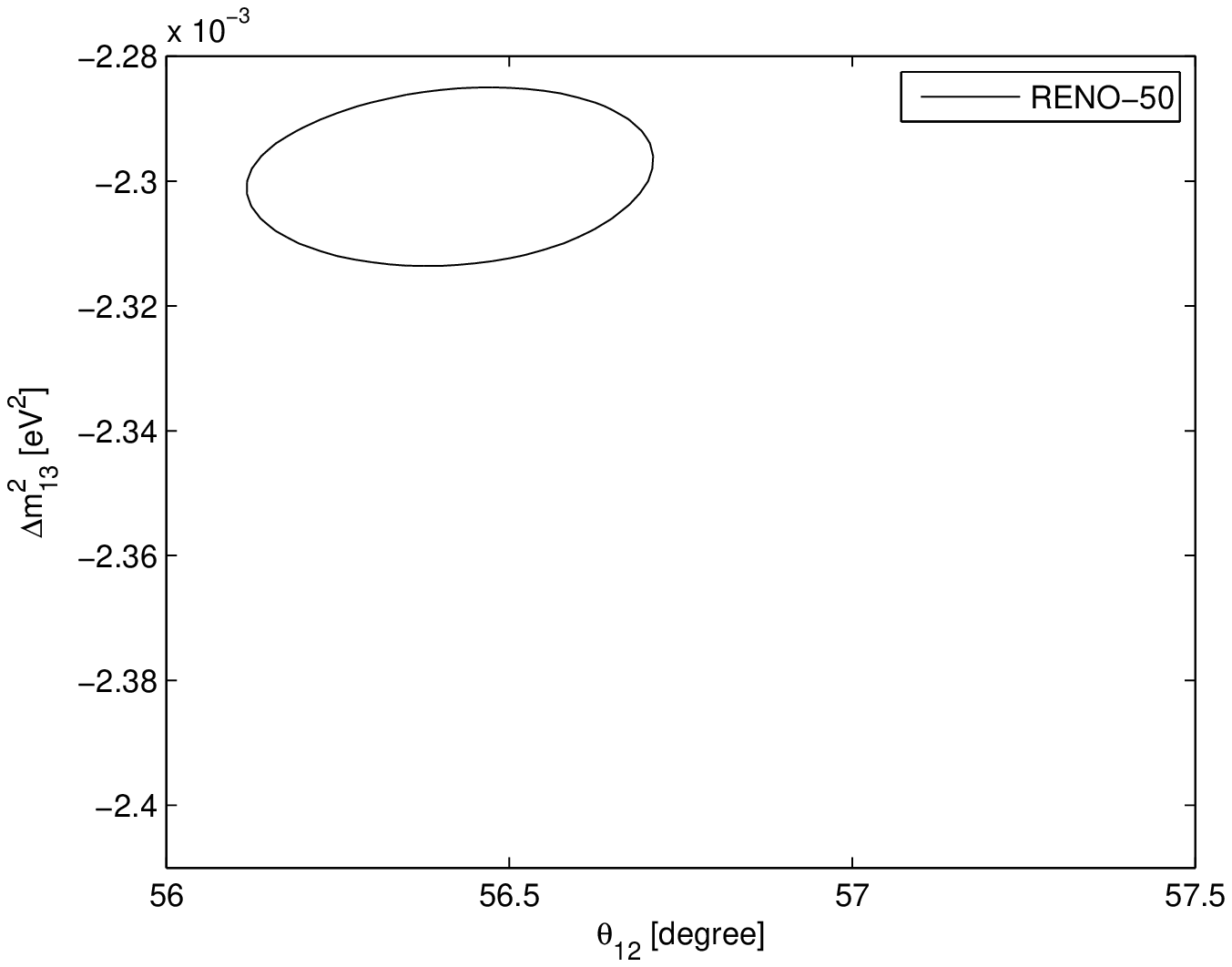}}
\end{center}
\vspace{2cm} \caption[]{The same as Fig. \ref{Contour-bright} except
that we have taken the true values to be $\Delta
m_{31}^2=2.417\times10^{-3}~{\rm eV}^2$ and
$\theta_{12}=56.43^\circ$. That is we have taken the LMA-dark
solution instead of the standard LMA solution.}

\label{Contour-dark}

\end{figure}

Figs (\ref{Contour-bright},\ref{Contour-dark}) show the  potential
of JUNO and RENO-50 experiments in determining both hierarchy and
sign($\cos 2
\theta_{12}$) after five years of data taking. 
We have assumed normal hierarchy and have taken the true value of
$\theta_{12}$ to be equal to $\theta_{12}= 33.57^\circ$ in Fig.
(\ref{Contour-bright}) and equal to $\theta_{12}= 56.43^\circ$ in Fig.
(\ref{Contour-dark}).  Contours show the 3 $\sigma$ C.L. solutions.
Notice that the determination of $|\Delta m_{31}^2|$ by either of
these  experiments will be far more precise than what is obtained by
global analysis of the present data both  in the absence of NSI
\cite{GonzalezGarcia:2012sz} and in its presence \cite{Maltoni}.
They can also remarkably improve the precision on $\theta_{12}$.
After five years of data taking, the precision of $\theta_{12}$ will
reach a remarkable value of $\Delta \theta_{12} =\pm 0.4^\circ$ or
better at 3$\sigma$ C.L. For ruling out the wrong hierarchy, we have
checked our result against that in Ref. \cite{Blen} and it seems our
results are in agreement.

From Fig (\ref{Contour-bright}-a) and Fig. (\ref{Contour-dark}-b),
we observe that  JUNO  can determine these parameters more precisely
than  RENO-50 would. This is mainly due to the fact that the reactor
power and therefore neutrino flux are higher at JUNO. As seen from
Figs. (\ref{Contour-dark}-a and -d), while at 3 $\sigma$ RENO-50
finds solutions with wrong sign($\Delta m_{31}^2$) or wrong
sign($\cos 2\theta_{12}$), JUNO rules out these wrong solutions. We
have found that when LMA-Dark is taken as the true solution,
RENO-50, JUNO and their combined results rule out the wrong LMA
solution with $\chi^2=5.5$ ({\it i.e.,} $>90 ~\%$ C.L.),
$\chi^2=12.9$ ({\it i.e.,} $\sim 3 \sigma$ C.L.) and $\chi^2=19.94$
({\it i.e.,} $\sim 4 \sigma$ C.L.), respectively.
 Similarly for
standard LMA solution with $\cos 2\theta_{12}>0$, RENO-50, JUNO and
their combined results rule out the wrong LMA-Dark solution with
$\chi^2=4.95$ ({\it i.e.,} $>90 ~\%$ C.L.), $\chi^2=11.4$ ({\it
i.e.,} slightly less than $ 3 \sigma$ C.L.) and $\chi^2=18.34$ ({\it
i.e.,} slightly less than $4 \sigma$ C.L.), respectively. Turning
off the background, JUNO can also rule out the wrong LMA-Dark
solution at more than $3\sigma$ C.L.
From Figs. (\ref{Contour-bright}) and (\ref{Contour-dark}), we also
see that the precision by JUNO is overally better. Remember that we
had assumed similar calibration uncertainty, energy resolution and
background for these two experiments. By varying the calibration
error by a factor of two we have found that the results from these
two setup do not change much. However, as expected, similarly to the
case of hierarchy determination \cite{ematters,Ciuffoli} the results
are very sensitive to the energy resolution. For example, if we
change the energy resolution from $3\%~ (E_\nu/{\rm MeV})^{1/2}$ to
$3.5\% ~(E_\nu/{\rm MeV})^{1/2}$, the wrong solution becomes
acceptable at 3 $\sigma$ C.L. by combined five years data  of JUNO
and RENO-50.

 As seen from the Figs. (\ref{Contour-bright}), the reactor experiments
cannot distinguish between the solution with $\cos 2\theta_{12}>0$
and $\Delta m_{31}^2>0$ and the one with $\cos 2\theta_{12}<0$ and
$\Delta m_{31}^2<0$. This degeneracy  is the result of the symmetry
under transformations in (\ref{transform}) when matter effects are
neglected. The subdominant matter effects slightly lift this
degeneracy but not enough to render them distinguishable. In the next section, we discuss whether
alternative methods to determine sign($\Delta m_{31}^2$) based on
matter effects by long baseline experiments or atmospheric neutrino
experiments can lift this degeneracy. The LMA-Dark solution
can be tested by neutrino  scattering experiments sensitive to NSI
effect. Similar discussion can be repeated for Fig.
(\ref{Contour-dark})
 where the LMA-Dark solution is taken as the true
solution.


 A similar
discussion also applies for inverted hierarchy: Contours for
inverted hierarchy with $\cos (2\theta_{12})>0$ and $\cos
(2\theta_{12})<0$ are very similar  respectively  to Fig.
(\ref{Contour-dark}) and Fig. (\ref{Contour-bright}).
\section{Degeneracy  and matter effects\label{degeneracy}}
In this section, we generalize the symmetry under transformation shown in Eq (\ref{transform}) to all oscillation modes taking into account the matter effects on oscillation. A similar approach is also taken  in \cite{Maltoni}. The effective Hamiltonian
governing the evolution of neutrino states in the presence of matter effects can be written as
\begin{equation} H
 =V_{vacc}+V_{eff} \ \ {\rm where} \ \ V_{vacc}=U_{PMNS}\cdot Diag(\Delta_1,\Delta_2,\Delta_3) \cdot U_{PMNS}^{T} ,\end{equation}
 in which $\Delta_i=m_i^2/(2E_\nu)$ and
 \begin{eqnarray} U_{PMNS}=\left[ \begin{matrix}  c_{12}c_{13} & s_{12} c_{13} & s_{13} e^{i \delta} \cr
 -s_{12}c_{23}-c_{12}s_{23}s_{13} e^{i\delta} & c_{12} c_{23} -s_{12} s_{23} s_{13}e^{i\delta} & s_{23} c_{13} \cr
  s_{12}s_{23}-c_{12}c_{23}s_{13} e^{i\delta} & -c_{12} s_{23} -s_{12} c_{23} s_{13}e^{i\delta} & c_{23} c_{13}
  \end{matrix} \right] .
  \end{eqnarray}
  $V_{eff}$ is a matrix describing both standard and non-standard matter effects. For the standard case $V_{eff}$ is diagonal with $(V_{eff})_{\mu \mu}=(V_{eff})_{\tau \tau}$.
Replacing $\theta_{12} \to \pi/2 -\theta_{12}$, $\delta\to \delta+\pi$ and $\Delta_1 \leftrightarrow \Delta_2$, $V_{vacc}$ will transform into $ S\cdot V_{vacc}\cdot S$ where $S=Diag(1,-1,-1)$. Since we have the freedom of rephasing $\nu_\alpha$, the oscillation probabilities will remain the same provided that
at the same time, $V_{eff} \to S\cdot V_{eff} \cdot S$; {\it i.e.,} $(V_{eff})_{ e\mu} \to - (V_{eff})_{ e\mu}$ and $(V_{eff})_{ e\tau} \to - (V_{eff})_{ e\tau}$.
Replacing $\Delta_1 \leftrightarrow \Delta_2$ is equivalent to $\Delta_{21} \to - \Delta_{21}$ and $\Delta_{31} \to \Delta_{31}-\Delta_{21}$. On the other hand the evolutions with $H$
 and $-H^*$ lead to the same oscillation probabilities \cite{Maltoni}. Thus, the oscillation probability will be the same if we simultaneously replace
 \be \label{tran-gen} \theta_{12} \to \frac{\pi}{2}-\theta_{12}, \ \ \delta \to \pi -\delta, \ \ \Delta_{31} \to -\Delta_{13}+\Delta_{21} \ \ {\rm and} \ \  V_{eff} \to - S\cdot V_{eff} \cdot S . \ee

 Notice that the transformation in Eq. (\ref{transform}) is a subset of these transformations. Since for reactor neutrinos, $\delta$ and matter effects ($V_{eff}$) are irrelevant, we did not need to include the transformations of $\delta$ and $V_{eff}$ in Eq. (\ref{transform}). Within the SM, $V_{eff}$ is fixed by the composition of the medium and the Fermi constant: $(V_{eff})_{ee}=\sqrt{2} G_F N_e-\sqrt{2} G_F N_n/2 $ and $(V_{eff})_{\mu \mu}=-\sqrt{2} G_F N_n/2 $.  As a result, replacing $V_{eff} \to -S\cdot V_{eff}\cdot S$ is meaningless. However in presence of NSI for a given matter composition, such transformation can be interpreted as shifts in values of $\epsilon_{\alpha \beta}$ which parameterizes new physics. Following \cite{Maltoni}, let us focus on NSI with $u$- and $d$-quarks parameterized respectively by
 $\epsilon_{\alpha \beta}^u$ and $\epsilon_{\beta \alpha}^d$. The effect of NSI on neutrino oscillation in an electrically neutral medium is described \cite{Maltoni} by
 $$\epsilon_{\alpha\beta}=Y_u  \epsilon_{\alpha\beta}^u+  Y_d  \epsilon_{\alpha\beta}^d$$
 where $Y_u=2+Y_n$ and $Y_d=1+2Y_n$ in which $Y_n$ is the neutron to electron ratio. For Long baseline experiments, $Y_n=1.012$ \cite{Maltoni,Prem}. The fact that the $Y_n$ composition of the Sun and Earth are different can help us to partially solve the degeneracy.

 Ref. \cite{Maltoni} has made a global analysis of data and has found that at 3$\sigma$ C.L., the allowed range of $\epsilon$ for the LMA solution
 with $\cos 2\theta_{12}>0$ is  \begin{equation} \label{ran1}-0.6<\epsilon_{ee}-\epsilon_{\mu \mu}<4 \end{equation}
 and for the LMA-Dark solution with  $\cos 2\theta_{12}<0$, the allowed range is
  \begin{equation} \label{ran2}-8<\epsilon_{ee}-\epsilon_{\mu \mu}<-4 .\end{equation}
 As expected, while the LMA-dark solution requires $
\epsilon \ne 0$, the LMA solution includes $\epsilon=0$.  Without loss of generality we can set $\epsilon_{\mu \mu}=0$  because subtracting a matrix  proportional to unit matrix ({\it e.g.,} $(V_{eff})_{\mu\mu} I$) from $H$ will not affect the oscillation probabilities. With this convention, $V_{eff} \to -S\cdot V_{eff} \cdot S$ corresponds to
$$\epsilon_{ee}+1 \to -(1+ \epsilon_{ee}).$$
Symmetry under transformation in Eq. (\ref{tran-gen}) therefore implies that the part of  LMA solution with $2<\epsilon<4$ cannot be distinguished from LMA-Dark solution with $-4<\epsilon<-6$ and opposite hierarchy by oscillation experiments taking place in the earth ({\it i.e.,} by  reactor, atmospheric and long baseline experiments). However, the rest of the range in Eqs. (\ref{ran1}) and (\ref{ran2}) can be in principle distinguished by long baseline and atmospheric neutrino experiments sensitive to matter effects on oscillation.

We examined the possibility of solving degeneracy by using the NOvA experiment. Sensitivity of NOvA to NSI had also been discussed in \cite{Shoe}.
 We used the GLoBES software to carry out the analysis. {Details of the simulation of NOvA experiment is based on \cite{Ayres:2004js,nova}.}
For true values we have taken $\theta_{12}={33.57}$ and  set all the NSI parameters to zero;
 $\epsilon=0$.   We have assumed normal hierarchical scheme.  We have found that after six years of data taking ({\it i.e.,} 3 years in neutrino mode and 3 years in antineutrino mode), NOvA can rule out the other solution with opposite sign of $\cos 2 \theta_{12}$ and $\Delta_{31}$ with $\chi^2= 3.9$ which for 2 dof corresponds to about 85\% C.L.

\section{Conclusions \label{Con}}

We have examined the potential of the intermediate baseline reactor
experiments in  discriminating between LMA and LMA-Dark solutions.
This method is based on determining sign($\cos 2\theta_{12}$) rather
than probing the NSI. Sensitivity to sign($\cos 2\theta_{12}$) ({\it
i.e.,} LMA versus LMA-Dark solutions) as well as to sign($\Delta
m_{31}^2$) ({\it i.e.,} normal versus inverted mass ordering) both
appear in oscillatory terms  in the survival probability,
$P(\bar{\nu}_e \to \bar{\nu}_e)$ that are given by $\Delta m_{31}^2$
and are suppressed by $s_{13}^2$. Thus, to disentangle their
effects, the following challenges have to be overcome: (1) the
statistics should be high enough; (2) the energy resolution, $\delta
E_\nu/E_\nu$, should be  small enough to resolve the oscillatory
terms given by $(\Delta m_{31}^2 L/E_\nu)$ and (3) the effects of
oscillatory terms given by $\Delta m_{31}^2$ should not be washed
out by averaging over  baselines of various reactor cores
contributing to the flux. These conditions will be fulfilled at the
JUNO and RENO-50 experiments. We have found that for a given hierarchy RENO-50,  JUNO and
combined RENO-50 and JUNO results can discriminate between LMA and
LMA-Dark solution, respectively, at $>90 ~\%$ C.L.,  $\sim 3
\sigma$ C.L. and $\sim 4 \sigma$ C.L. after five years.

We have demonstrated that neglecting the matter effects,
$P(\bar{\nu}_e \to \bar{\nu}_e)$ becomes symmetric under
transformation in Eq.~(\ref{transform}). This means there is a
degeneracy between solutions for which both the mass hierarchy and
the sign of $\cos 2 \theta_{12}$ are simultaneously flipped. Matter
effects can to some extent lift this degeneracy but not enough in
order for JUNO and RENO-50  to resolve this degeneracy. Moreover, when we allow a shift in values of NSI parameters, the symmetry can be generalized to include matter effects as described in Eq. (\ref{tran-gen}). The degeneracy can be partially solved by combining data from long baseline experiments sensitive to matter effects  and the solar neutrino data thanks to the fact  that the medium in the Sun and in the Earth have different compositions {\it i.e.,} neutron to electron ratio. In particular, we found that after six years of data taking, the NOvA experiment can discriminate  between the LMA solutions with $\cos 2\theta_{12}>0$ and no NSI ($\epsilon=0$) and the LMA-Dark solution with opposite mass ordering with about 85 \% C.L.
Moreover experiments probing neutral current NSI such as neutrino
scattering experiments can test LMA-Dark solution and hence break
this degeneracy.

 \section*  {Acknowledgements}
 The authors are grateful to T. Schwetz, J. Evslin and M. Maltoni for useful remarks.
 Y.F.  acknowledges partial support from the  European Union FP7  ITN INVISIBLES (Marie Curie Actions, PITN- GA-2011- 289442).


\end{document}